\documentclass[useAMS,usenatbib]{mn2e}

\usepackage{psfig}

\title[A detection of the donor star of Aql\,X-1?]{A detection of the donor star of Aquila X-1 during its 2004 outburst?}
\author[R. Cornelisse et~al.]{R. Cornelisse$^{1,2}$\thanks{E-mail:
corneli@iac.es}, J. Casares$^{2}$, D. Steeghs$^{3}$,  A.D. Barnes$^{1}$, P.A. Charles$^{4,1}$, 
\newauthor
R.I. Hynes$^{5}$, K. O'Brien$^{6}$\\
$^{1}$School of Physics and Astronomy, University of Southampton, Highfield, Southampton SO17 1BJ, UK\\
$^{2}$Instituto de Astrofisica de Canarias, Via Lactea, La Laguna E-38200, Santa Cruz de Tenerife, Spain\\
$^{3}$Harvard-Smithsonian Center for Astrophysics, 60 Garden Street, Cambridge, MA 02138, USA\\
$^{4}$ South Africa Astronomical Observatory, P.O.Box 9.Observatory 7935, South Africa\\
$^{5}$Department of Physics and Astronomy, 202 Nicholson Hall, Louisiana State University, Baton Rouge, LA 70803, USA\\
$^{6}$European Southern Observatory, Casilla 19001, Santiago 19, Chile\\
}

\begin{document}

\date{Accepted  Received ; in original form }

\pagerange{\pageref{firstpage}--\pageref{lastpage}} \pubyear{2004}

\maketitle

\label{firstpage}

\begin{abstract}
Phase-resolved high resolution optical spectroscopy has revealed
narrow N\,III and He\,II emission lines from the soft X-ray transient
Aquila\,X-1 during its 2004 outburst that move as a function of the
orbit consistent with the phasing of the donor star. Under the
assumption that these lines come from the irradiated side of the donor
star, we can constrain its $K_2$ velocity to $\ge$247$\pm$8 km s$^{-1}$,
and derive a mass function of $f(M_1)$$\ge$1.23$\pm$0.12 $M_\odot$.
Estimates for the rotational broadening based on the emission
components suggest a possible massive neutron star of
$\ge$1.6$M_\odot$ (at 95\% confidence). However, an updated
ephemeris and additional high resolution spectroscopy of Aql\,X-1
during a future outburst are warranted in order to confirm that the
narrow lines indeed originate on the donor star surface, and reliably
characterise the system parameters of this important X-ray binary.
Spectra taken during the end of the outburst show that
the morphology of the emission lines changed dramatically. No donor
star signature was present anymore, while the presence of narrow
low-velocity emission lines became clear, making Aql\,X-1 a
member of the slowly growing class of low-velocity emission line
sources.
\end{abstract}


\section{Introduction}

Low mass X-ray binaries (LMXBs) are systems where a compact object,
either a neutron star or black hole, accretes material via Roche-lobe
overflow from a low mass ($\le$1$M_\odot$) companion. Soft X-ray
transients (SXTs) are an important sub-class of the LMXBs. Contrary to
the persistent LMXBs, that always accrete material at high rates, they
occasionally have short outbursts ($\simeq$weeks up to months) before
returning back to quiescence. These systems are especially interesting for the
prospect of quiescent studies of the mass donor, since they provide
the only direct method to obtain orbital solutions and system masses
for LMXBs (e.g. Charles \& Coe 2005).  Such system parameter studies
are usually not possible during an outburst because the photospheric
emission from the companion star is swamped by the emission from the
X-ray irradiated disk. Unfortunately, in quiescence many SXT companion
stars are too faint ($V$$>$22-23) to allow kinematic studies. This has
substantially hampered detailed interpretation of their evolution, and
our ability to construct compact object mass distributions.

Aql\,X-1 is a SXT that shows one of the shortest
recurrence times, making it one of the most intensively studied
systems. Aql\,X-1 shows quasi-periodic outbursts, with approximately one
year intervals (Priedhorsky \& Terrell 1984). During such an outburst
it shows Type\,I X-ray bursts, indicating that the compact object is a
neutron star (Koyama et~al. 1981; Czerny et~al. 1987). Its optical
counterpart was detected by Thorstensen et~al. (1978) during an
outburst at a magnitude of $B$$\simeq$17. An 18.97 hr periodic
modulation in the optical was observed during an outburst, which was
interpreted as the orbital period (Chevalier \& Ilovaisky 1991). In
quiescence the counterpart fades to $V$$\simeq$21.6, but is only
0.46$''$ away from a $V$=19.26 star (Chevalier et~al. 1999), severely
hampering kinematic studies of Aql\,X-1. Despite intense studies there
are currently no strong constraints on the system parameters.

Photometric observations show that the light curve is double humped
during quiescence, changing to single humped in outburst (e.g. Welsh
et~al. 2000; Chevalier \& Ilovaisky 1991,1998; Shahbaz
et~al. 1998). This is interpreted as ellipsoidal variations from the
Roche-lobe filling donor star in quiescence, changing in outburst to
reflection due to heating of the side of the secondary facing the
neutron star. This suggests that during an outburst the optical light
is dominated by the irradiated side of the secondary (Welsh
et~al. 2000).  The spectrum of Aql\,X-1 in quiescence, after careful
deconvolution of the spectrum of the contamination star, shows
absorption features that can be associated with a K6-M0 V star, as
well as emission lines typical for LMXBs (Chevalier et~al. 1999).

In this paper we will explore the possibility to determine the system
parameters of Aql\,X-1 during an outburst using narrow Bowen
fluorescence lines. This technique was first applied to Sco\,X-1
(Steeghs \& Casares 2002). Narrow emission line components
coming from the irradiated companion star, especially in the Bowen
blend (a blend of N\,III 4634/4640 \AA\ and C\,III 4647/4650 \AA\
lines), were used to trace its orbital motion and determine the mass
function. Since then, this technique has led to the discovery of donor
star signatures in other persistent LMXBs such as X\,1822$-$371,
4U\,1636$-$53 and 4U\,1735$-$44 (Casares et~al. 2003, Casares
et~al. 2006), or transient sources where the optical counterpart is
too faint in quiescence, such as GX\,339$-$4 (Hynes et~al. 2003a).

\begin{figure}
\psfig{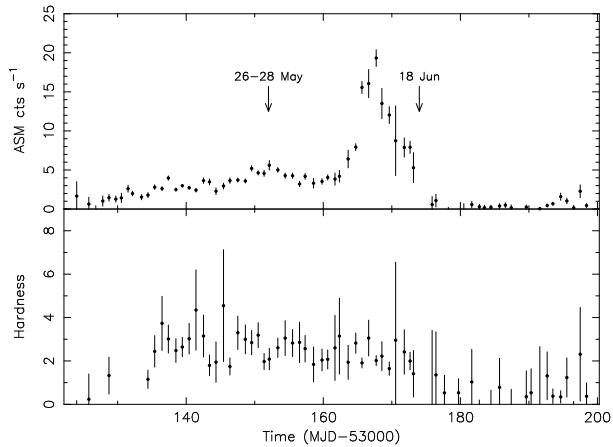}
\caption{Top: X-ray light curve of Aql\,X-1 obtained by the All Sky
Monitor onboard RXTE showing the 2004 outburst. The arrows
indicate the times of our optical spectroscopy. Bottom: X-ray hardness
ratio defined as 3.0-12.1 keV/1.3-3.0 keV.
\label{asm}}
\end{figure}

\section{observations and data reduction}

From May until June 2004 Aql\,X-1 was in outburst (see
Fig.\,\ref{asm}), and we were able to obtain a total of 28 spectra
using the FORS-2 spectrograph attached to the VLT/UT4 (Yepun
Telescope) at Paranal Observatory (ESO). In Table\,\ref{log} we give
an overview of all the observations.  The spectra were taken with the
1400V volume-phased holographic grism, and a slit width of 0.7$''$,
giving a wavelength coverage of 4514-5815 \AA with a
resolution of 70 km s$^{-1}$ (FWHM). The seeing during these
observations varied between 0.5 and 2.1 arcsec. During daytime He, Ne,
Hg and Cd arc lamp exposures were taken for the wavelength calibration
scale.

\begin{table}\begin{center}
\caption{Overview of the observations of Aql\,X-1. Indicated are the
UT dates of each observation, $R$ magnitude according to observations by
Maitra \& Bailyn (2005), number of observations during that night, 
exposure time, and corresponding orbital phases of the
observations according to the weighted ephemerides by Chevalier et~al. (1991,1998) and Welsh et~al. (2000).
\label{log}}
\begin{tabular}{ccccc}
\hline
Date      & $R$ & No. & Exp. & Orb. phases\\
(dd-mm-yy)& (mag)     &           & (s)  & \\
\hline
26-05-04 & 16.3 & 1  & 900  & 0.91\\
27-05-04 & 16.3 & 8  & 600  & 0.88-0.90,0.14-0.17\\
28-05-04 & 16.3 & 11        & 900  & 0.17, 0.30-0.43\\
18-06-04 & 17.5 & 8         & 674  & 0.76-0.84\\
\hline
\end{tabular}
\end{center}\end{table}

We de-biased and flat-fielded all the images and used optimal
extraction techniques to maximise the signal-to-noise ratio of the
extracted spectra (Horne 1986). The pixel-to-wavelength scale was
determined using a 4th-order polynomial fit to 20 reference lines
resulting in a dispersion of 0.64 \AA pix$^{-1}$ and rms scatter
$<$0.05 \AA. We also corrected for velocity drifts due to
instrumental flexure (always $<$5 km s$^{-1}$) by cross-correlating
the sky spectra using a small region around the [O\,I] sky line
at $\lambda$5577.4. We thus realigned all spectra using this single 
skyline as the zero point.

\section{Data analysis}

\begin{figure*}
\psfig{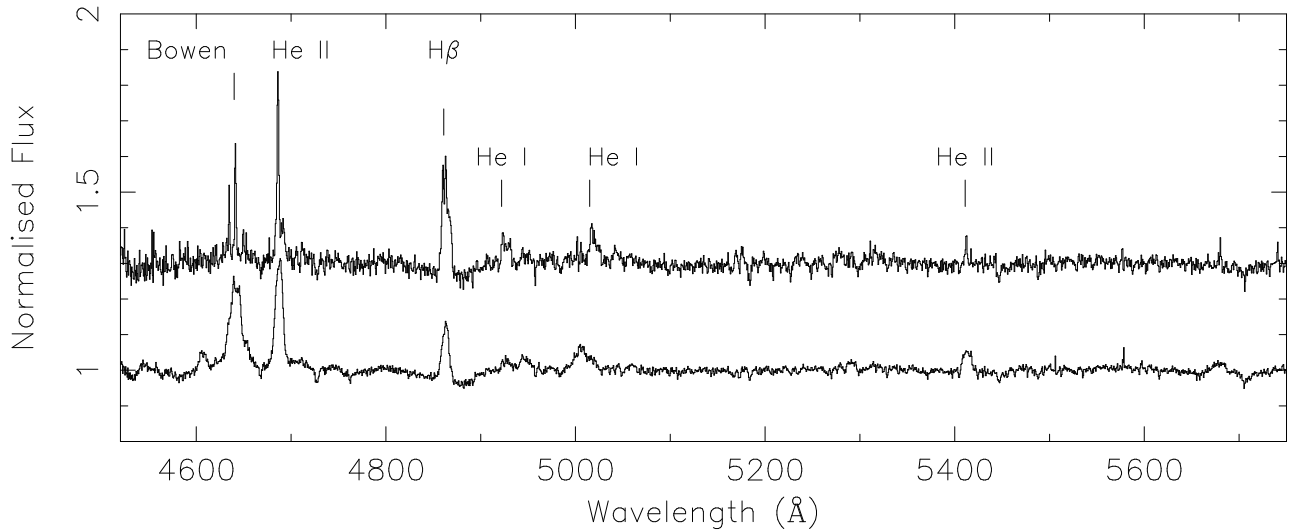}
\caption{Average optical spectrum of Aql\,X-1 during 26 to 28 May 2004
 run (bottom), and during the 18 June observations (top). We
have indicated the most prominent lines.
\label{spectrum}}
\end{figure*}

\begin{figure}\begin{center}
\psfig{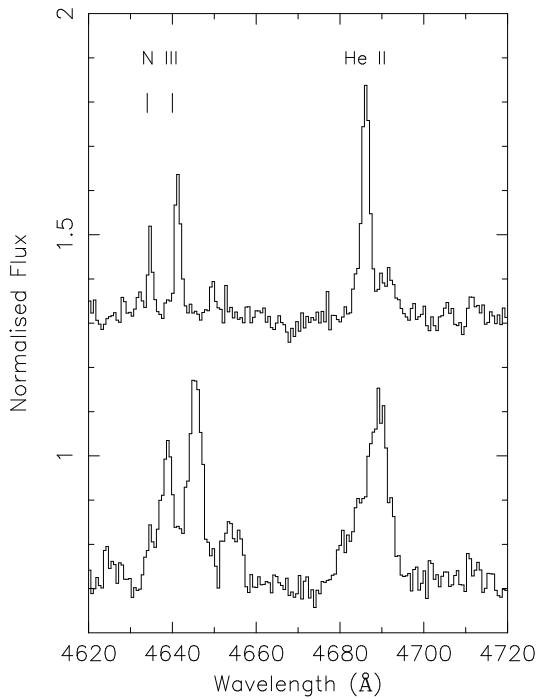}
\caption{Close-up of the He\,II $\lambda4686$ and Bowen region for a
single observation obtained on 27 May 2004 (bottom) and an average of
all observations taken on 18 June 2004 (top).
\label{bowen}}
\end{center}\end{figure}

\begin{figure}
\psfig{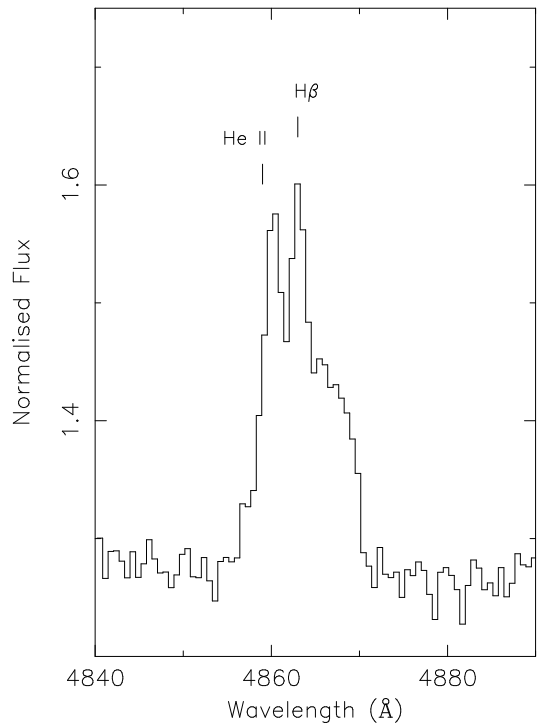}
\caption{Close-up of H$\beta$ during the June 2004 observations.
\label{hbeta}}
\end{figure}

Fig.\,\ref{spectrum} shows the average spectrum of both the 26-28 May
2004 run (bottom) and the 18 June 2004 observations (top).  We note
that the spectra are dominated by high excitation emission lines,
notably from He\,II $\lambda$4686, He\,II $\lambda$5411, H$\beta$ and
the Bowen region $\lambda$$\lambda$4630-4650, as expected from an
X-ray binary during outburst.  Comparing the two spectra in
Fig.\,\ref{spectrum}, and especially the close-ups of the most
important emission lines in Figs.\,\ref{bowen}\&\ref{hbeta}, the
first thing we notice is that the morphology of the emission lines has
changed significantly. During the 26-28 May 2004 run the emission
lines consist of a broad component (most likely coming from the
accretion disk), and a narrow component that shows a significant
velocity shift between the individual spectra (these components are
clearest in Fig.\,\ref{bowen}). On the other hand, during the 18 June
2004 observations the lines are exceptionally narrow (see
Table\,\ref{lines}), and both the broad base and the moving component
have disappeared. In order to quantify the change in emission
lines, we have listed the equivalent widths, ratio of peak flux over
continuum flux and FWHM of the main lines derived from average spectra
of the May and June 2004 observations in Table\,\ref{lines}.
Furthermore, we have also indicated the $R$ magnitude, as determined
by Maitra \& Bailyn (2005), in Table\,\ref{log}. The only line that
has broadened a little compared to previous observations is H$\beta$.
However, we do note that the He\,II $\lambda$4859 emission line from
the Pickering series is also present (see Fig.\,\ref{hbeta}), and
could contribute to this broadening, as was also the case in the LMXB
XTE\,J2123$-$058 (Hynes et~al. 2001). Most importantly, the narrow
lines in the June data do not show any orbital motion, despite phase
coverage of $\simeq$0.1$P_{\rm orb}$ near the orbital conjunction.  A
close inspection of the individual May 2004 spectra shows that these
weak and narrow stationary lines are also present but swamped by the
much stronger disk and moving components. Although these narrow lines
are obscured most of the time by the moving lines, the data is
consistent with them being present in all spectra. Unfortunately,
since we do not have an absolute flux calibration for the spectra, it
is not clear if the line flux from these stationary components has
evolved between May and June 2004.

\begin{figure}\begin{center}
\psfig{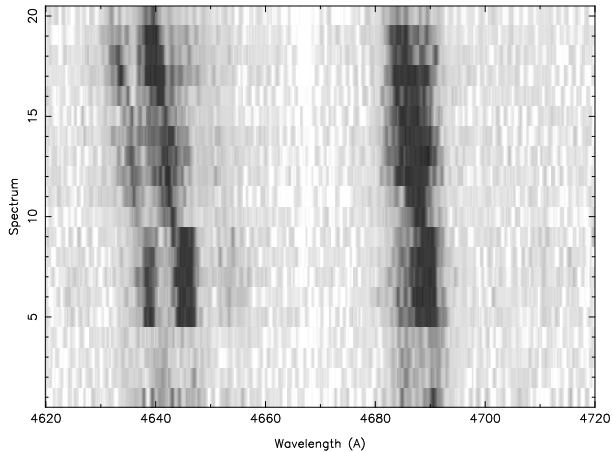}
\caption{Trail of the Bowen and He\,II $\lambda$4686 lines clearly
showing the movement of the lines due to orbital motion. Note that we
have not phase-binned the spectra due to the sparse sampling.
\label{trail}}
\end{center}\end{figure}

Taking a closer look at the region around the Bowen
$\lambda$$\lambda$4630-4650 emission in the individual spectra (see
Fig.\,\ref{bowen}), we notice that it is dominated by narrow emission
lines from N\,III $\lambda$4640/4634. Curiously, there is no
indication of the presence of narrow C\,III lines , as was the case in
e.g. Sco\,X-1 (Steeghs \& Casares 2002), but we do note that also in
other LMXBs the N\,III Bowen lines are stronger than the C\,III
lines (e.g. Casares et~al. 2006). Fig.\,\ref{trail} shows that these
N\,III lines are moving, and we interpret this as orbital motion of a
confined region in the system. Furthermore, a close inspection of
He\,II $\lambda$4686 shows that there is a strong narrow component
(making up $\simeq$30\% of the total line flux) that appears to be
moving in phase with the narrow Bowen lines (see Fig.\,\ref{trail}).
Given that the off-set between the narrow He\,II and Bowen lines does
not change between the spectra strengthens our identification of the
narrow Bowen lines as N\,III.

In order to estimate the orbital phases of our spectra we started
with the orbital period derived by Chevalier et~al. (1998) and phase
zero by Welsh et~al. (2000). Phase zero is defined as the minimum in
the photometric lightcurve, which is expected to correspond to
inferior conjunction of the secondary if the photometric lightcurve is
indeed dominated by its emission. Unfortunately, this phasing does not
match up with the published phase zero by Chevalier et~al. (1991). The
differences between these zero points is 0.1 orbital phase, while the
error suggests only a difference of 0.04 orbital phase. Since we
cannot be certain what causes this discrepancy, we used a weighted
average of both ephemerides (assuming all quoted errors are correct)
to derive the orbital phase for our spectra. We show the results in
Table\,\ref{log}. In order to estimate the phase uncertainty, we use a
combination of the weighted formal error of two ephemerides (0.05
orbital phase) plus a systematic error of 0.09 orbital phase derived
from the difference between the formal and observed error between the
2 ephemerides. This gives a total minimum phase uncertainty on our
spectra of 0.1 orbital phase.

\begin{table}\begin{center}
\caption{Overview of the characteristics of the average main emission
lines observed in Aql\,X-1 during the May and June 2004 observations
respectively. For each line the equivalent widths (EW), ratio of peak
($F_{\rm peak}$) over continuum flux ($F_{\rm cont}$), and {\it FWHM}
are shown. \label{lines}}
\begin{tabular}{llcc}
\hline
Line & & May 2004 & June 2004\\
\hline
H$\beta$ \\
 &EW (\AA)                     & 0.52$\pm$0.01 & 1.47$\pm$0.04\\
 &$F_{\rm peak}$/$F_{\rm con}$ & 1.2$\pm$0.1   & 1.4$\pm$0.1\\
 & FWHM (km s$^{-1}$)          & 447$\pm$6     & 561$\pm$13\\
\hline
He\,II \\
 &EW (\AA)                     & 3.49$\pm$0.02 & 2.26$\pm$0.06\\
 &$F_{\rm peak}$/$F_{\rm con}$ & 1.3$\pm$0.1   & 1.6$\pm$0.1\\
 & FWHM (km s$^{-1}$)          & 539$\pm$5     & 153$\pm$6\\
\hline
Bowen \\ 
 &EW (\AA)                     & 6.26$\pm$0.03 & 4.06$\pm$0.07\\
 &$F_{\rm peak}$/$F_{\rm con}$ & 1.3$\pm$0.1   & 1.3$\pm$0.1\\
 & FWHM (km s$^{-1}$)          & 919$\pm$11    & 99$\pm$6$^1$\\
\hline
\multicolumn{4}{l}{$^1$ For a single narrow Bowen line only.}
\end{tabular}
\end{center}\end{table}

 In order to determine the radial velocity for each spectrum we fitted
Gaussians to the two narrow N\,III lines plus narrow component of
He\,II $\lambda$4686 simultaneously, where the rest-wavelength of
each Gaussian and their separation were fixed. We then determined the
off-set from their rest-wavelength using a minimum $\chi^2$
technique. We also determined the off-set using each line
individually, but the differences are negligible. However, the
presence of the narrow stationary lines could have an influence on our
estimated off-sets, especially close to orbital phases 0.5 and 1.0
where they blend with the moving lines.

Fig.\,\ref{comp} shows our radial velocity measurements as a function
of orbital phase. Note that due to observing conditions we have
averaged the 4 spectra around orbital phase 0.9 into two bins to
increase the signal to noise. For illustration purposes we also
included the 18 June spectra (indicated with filled circles and 
labeled as '18 Jun'), where we
used the narrow, stationary, lines to determine the off-set. Fitting a
sine curve to the measurements, with the orbital period taken as
fixed, it is clear that the 18 June lines do not originate from the
same source in the binary frame, and we therefore do not include them
in our radial velocity fits. From a fit through the resulting points
we derived a velocity semi-amplitude of $K_{\rm em}$=247$\pm$8 km
s$^{-1}$, a mean velocity of $\gamma$=30$\pm$10 km s$^{-1}$ and phase
zero of $\phi_0$=0.88$\pm$0.05.  We determined the errors on the fit
by scaling the errors on the individual radial velocities by a
factor of $\simeq$4 such that a $\chi_\nu^2$=1 is achieved.  Note
that phase zero was a free parameter, and does not include any
systematic uncertainty in the ephemeris used which is expected to be
of order $\simeq$0.1. We do notice a small jump in the radial
velocities around orbital phase 0.33, due to the narrow stationary
lines that are starting to blend with the wing of the moving
components, but excluding these points did not change the results of
our fit.

A close inspection of the broad component of He\,II $\lambda$4686
shows that it moves in anti-phase with the narrow component. However,
depending on the orbital phase either the wings or the core is
dominated by the narrow component. We have tried to subtract this
narrow component, but were not able to do this reliably and could
therefore not measure the radial velocities for the broad component
from this line. Instead, we investigated the other prominent emission
lines and noted that the only line where a narrow component is not the
dominating feature was H$\beta$. However, also for this line we have
to be cautious since there is some contamination from He\,II
$\lambda4859$ as shown in Fig.\,\ref{rest}. We applied the double
Gaussian technique to the wings of this line (Schneider \& Young
1980). As we moved away from the line core, we noted that the velocity
semi-amplitude was stable at $K_1$=68$\pm$5 km s$^{-1}$, the
zero-phase was at $\phi_0$=0.35$\pm$0.05 and the mean velocity around
$\gamma$=130$\pm$20 km s$^{-1}$ (while keeping the orbital period
fixed again). These values appear to be consistent with those from the
broad component in He\,II $\lambda$4686, and indeed show a 0.5 phase
difference between the broad and narrow components. However, we do
note that the difference in mean velocity between the narrow and broad
lines is consistent with the difference between H$\beta$ and He\,II
$\lambda$4859, suggesting that the latter is contributing
significantly to the blend. Furthermore, we also note that formally
the fit is not acceptable ($\chi^2$=458 for 36 d.o.f.), showing that
we must treat this result with caution.

\section{Donor Star Detection}

We have detected narrow N\,III fluorescence lines in the optical
spectrum of Aql\,X-1 during outburst that are very similar to the
donor star fluorescence features observed in other systems. Having
derived the radial velocity curve for these features, we will now
discuss the likelihood of these tracing the irradiated donor in
Aql\,X-1. These lines are a dominant feature of the Bowen blend, as
was also the case in GX\,339$-$4, Sco\,X-1, X\,1822$-$371, V801\,Ara
and V926\,Sco (Hynes et~al. 2003a, Steeghs \& Casares 2002, Casares
et~al. 2003+2006). In these X-ray binaries the narrow lines are
thought to originate on the surface of the companion star. This
is nicely illustrated in the eclipsing LMXB X\,1822-371. Here, the
origin of the narrow Bowen lines can unambiguously be claimed due to
the excellent agreement between the photometric and spectroscopic
ephemerides (Casares et~al. 2003). For the non-eclipsing LMXBs one has
to rely on ephemerides based on determining the minima from
lightcurves that are quite variable and far from sinusoidal. Again,
when the number of cycles is large enough one also finds an excellent
agreement with the spectroscopic ephemerides, as was the case for
V801\,Ara, and also explained the apparent difference between the
photometric and spectroscopic phasing in V926\,Sco (Casares
et~al. 2006). We therefore think that the most likely place of origin
of the narrow lines observed in Aql\,X-1 is also on the surface of the
companion star.

Unfortunately, due to the rather large uncertainty in the phasing (0.1
orbital phase), we cannot rule out the possibility that the narrow
lines in Aql\,X-1 do not originate on the companion star. However, for
the other two potential regions where these lines could originate, the
accretion stream or impact point, it is not clear if it could produce
such narrow emission lines as observed in Aql\,X-1. Comparing the
Doppler maps of several LMXBs for which a signature of these regions
is observed, namely EXO\,0748$-$678, X\,1822$-$371 and
XTE\,J1118$+$480, shows that the stream is observed either in
absorption or the region is too extended to produce such narrow lines
(Pearson et~al. 2006, Casares et~al. 2003, Torres
et~al. 2002). Furthermore, also in cataclysmic variables the accretion
stream or impact point are commonly observed, but again they are very
broad and do not tend to have sinusoidal radial velocity curves.  We
therefore think that the surface of the companion star is the most
likely place of origin for these N\,III narrow lines, although an
up-dated ephemeris and a Doppler map of the Bowen region are needed to
confirm this.

\subsection{Curious Features}

Narrow Bowen components have now been observed in growing list of X-ray
binaries (e.g. Steeghs \& Casares 2002, Casares et~al. 2003, Casares
et~al. 2006), and we can now also add Aql\,X-1 to
that list. Interestingly, the narrow lines in Aql\,X-1 show a strong
similarity with those observed in GX\,339$-$4, also an X-ray transient
but with a black hole primary that was confirmed by the observation of
the narrow Bowen components (Hynes et~al. 2003a).  In GX\,339$-$4 the
narrow components were of a comparable strength to those in Aql\,X-1 and
also dominate the Bowen region, while there was also an indication
that a narrow component was contaminating the He\,II emission (Hynes
et~al. 2003a).

From the Bowen/He\,II radial velocity curve we have estimated a
systemic velocity of 30$\pm$10 km s$^{-1}$, while H$\beta$ gives a
value of 130$\pm$20 km s$^{-1}$. This is most likely due to the
presence of He\,II $\lambda$4859, and would suggest that it is
dominating the blend. Such contamination has been observed in other
X-ray binaries such as XTE\,J2123$-$058 (Hynes et~al. 2001). However,
this does mean that we will have to be very careful with associating
this semi-amplitude with the orbital motion of the neutron star, since
this could also be distorted due to the blending of different
components. Such caution is further enhanced by the fact that
observations of, for example, X\,1822$-$371 by Cowley et~al.  (2003)
have shown that measuring the semi-amplitude of the compact object
from emission lines can only give a first order approximation.
A full orbital phase coverage of a
non-blended line that will not have a strong narrow component during a
future outburst of Aql\,X-1 will be needed to estimate
the $K$-velocity of the neutron star.

The most curious feature of Aql\,X-1 is the change of the emission
lines during the June 2004 observations. Such a dramatic change in the
morphology of the emission lines has been observed before in the X-ray
nova GRO\,J0422$+$32 (Shrader et~al. 1996). When the source hardened
in X-rays the He\,II $\lambda$4686 and the Bowen lines were suddenly
greatly reduced in intensity while H$\alpha$ and H$\beta$ became much
stronger. This change was interpreted as the ``turning
on'' of the hard X-ray component. Although, no hard X-ray/soft
gamma ray observations were performed for Aql\,X-1 around the time of
our observations in order to test this (Bird, private comm.), the
soft X-ray lightcurve (Fig.\,\ref{asm}) does indicate a large change in
the accretion activity between the May and June observations.

Even if such ``turning on'' of the hard X-ray component occurred in
Aql\,X-1 it is still strange that we do not observe the donor star
anymore in the June 2004 observations. Since these observations were
performed during the final stages of the outburst, i.e. a few days
before it went back into quiescence (Maitra \& Bailyn 2005), a large
fraction of the inner accretion disk might already have disappeared.
This could have stopped the production of UV photons that are needed
in the Bowen fluorescence process (McClintock et~al. 1975), and
thereby terminating the narrow lines coming from the companion.
Unfortunately, this scenario cannot be verified with the limited
spectral information that is contained in the ASM data. Another
suggestion could be that the opening angle of the accretion disk has
increased between the different observations, and is screening the
donor star at the end of the outburst (see also Sect.\,4.2).

\begin{figure}
\psfig{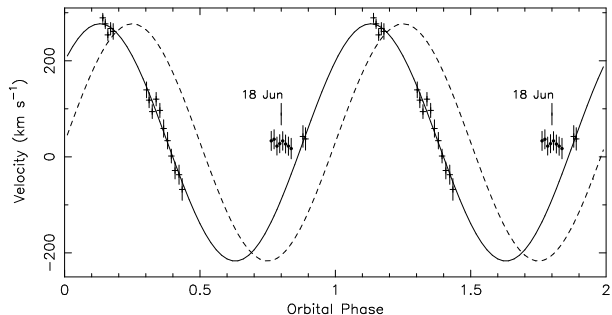}
\caption{Best fit radial velocity curve of the narrow Bowen blend
components and He\,II $\lambda$4686 as derived from a multi-Gaussian
fit. Phase zero corresponds to inferior conjunction of the donor star
according to the ephemeris derived in the text (and illustrated as the
dashed curve). Note that the error on the ephemeris is at least 0.1
orbital phase. Also indicated as filled circles and labeled '18 Jun'
are the radial velocities derived from the June 2004 spectra that have
been excluded from the fit. We have shown the curve twice for clarity.
\label{comp}}
\end{figure}

\begin{figure}
\psfig{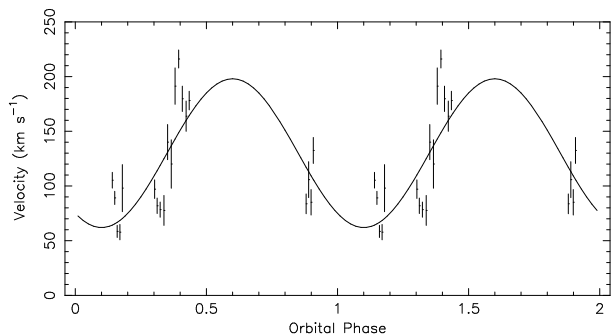}
\caption{Best fit radial velocity curve derived from H$\beta$ based on
the double Gaussian method. We have shown the curve twice for
clarity. 
\label{neutron}}
\end{figure}

The other curious feature of Aql\,X-1 is the narrow, stationary lines
that are present in both the May and June observations.  These narrow
lines do not show any change in radial velocity between the observed
spectra, and have a mean velocity of more or less 30 km s$^{-1}$,
which is interestingly comparable to the systemic velocity derived
from the lines that arise on the donor star. Aql\,X-1 is not the first
source where such low velocity emission lines are observed. For
example, in the dwarf novae IP Peg, SS Cyg and U Gem among others (see
Unda-Sanzana et~al.  2006 for an overview) also show such low velocity
emission lines (Steeghs et~al. 1996, Unda-Sunzana et~al. 2006). Also in the
LMXB Ser\,X-1 such unusually narrow and stationary lines were present
(Hynes et~al. 2003b). This makes Aql\,X-1 another LMXB to become a
member of the group that shows these low velocity emission lines, and clearly
indicates that these lines are a long-lived and common feature that
are independent of the spectral state of the source. Currently there
is no satisfactory explanation for these low velocity emission lines, but it is
becoming apparent that these lines are present in many compact
binaries. Therefore, a convincing explanation should be sought.
However, it is clear from Aql\,X-1 that they are not connected to the
companion, contrary to the narrow and moving lines observed in the May
2004 spectra.

\subsection{System parameters}

If we assume that we have detected a signature of the companion
star, what can we say, in combination with previous observations,
about the components of Aql\,X-1?  We know that the compact object is
a neutron star due to the presence of type\,I X-ray bursts (Koyama
et~al. 1981), leading to a mass of the compact object
$<$3.1$M_\odot$. Based on photometry and spectroscopy of the
optical counterpart of Aql\,X-1 (correcting for the interloper star at
a distance of 0.46 arcsec) Chevalier et~al. (1999) determined the
spectral type of the secondary as K6-M0.  Although several constraints
on the inclination exist, and even contradict each other (Shahbaz
et~al. 1997, 1998, Garcia et~al. 1999, Welsh et~al. 2000), most were
determined before it was known that the optical counterpart consisted
of a blend of two stars (Chevalier et~al. 1999). Shahbaz et~al. (1997)
determined an inclination of $\simeq$50$^\circ$ using the width of the
H$\alpha$ line, while Garcia et~al. (1999) argued that the inclination
must be $\simeq$70$^\circ$ based on the amplitude of the variations in
the outburst lightcurve.  Finally, Welsh et~al. (2000) used the
de-blended optical data to determine that the inclination is
$>$36$^\circ$, while the absence of dips or eclipses gives an
upper-limit on the inclination of $<$70-80$^\circ$ (Paczynski 1971).

\begin{figure}
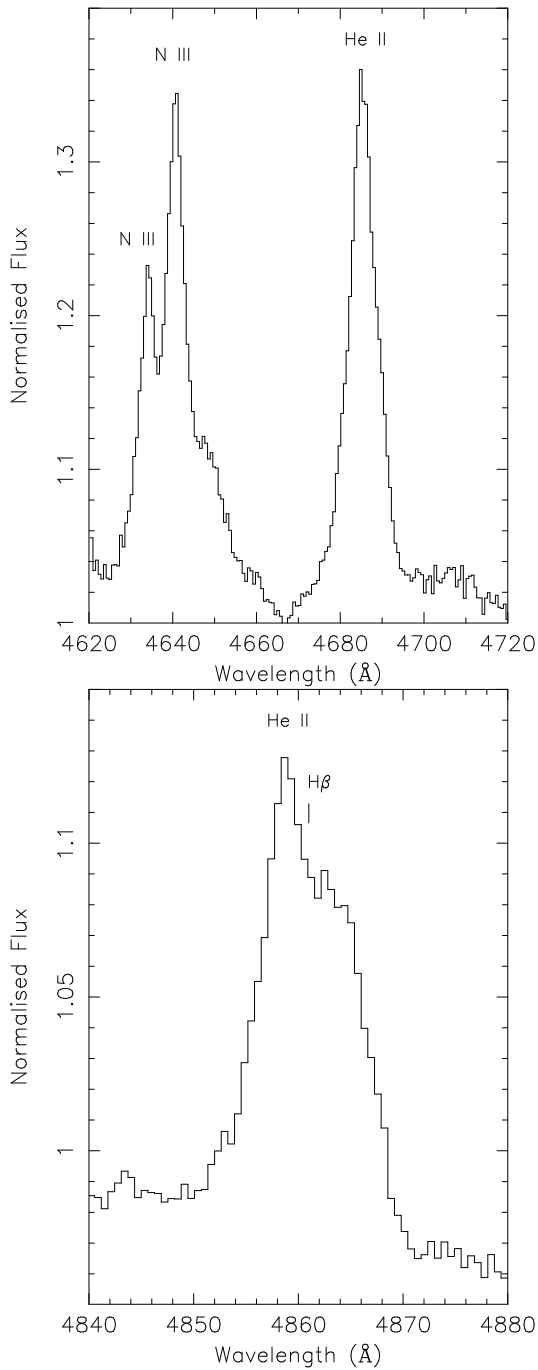
\begin{center}
\psfig{figure=aql_bowen_rest.ps,angle=-90,width=7.cm}
\psfig{figure=aql_hbeta_rest.ps,angle=-90,width=7.cm}
\caption{Top: Average spectrum of Aql\,X-1, around the He\,II
$\lambda$4686 and Bowen region, in the rest-frame of the donor
star. Bottom: Average spectrum of Aql\,X-1, around H$\beta$, in the
rest-frame of the donor star.
\label{rest}}
\end{center}\end{figure}

From Fig.\,\ref{comp} we have determined a $K_{\rm em}$ velocity of
247$\pm$8 km s$^{-1}$. However, since the narrow lines are expected to
arise on the surface of the donor star, this value will be a
lower-limit on the centre of mass velocity of the secondary. Without
using any other constraint on Aql\,X-1 we can determine that the mass
function $f(M_1)$=$M_1\sin^3i/(1+q)^2$$>$1.23$\pm$0.12$M_\odot$, only
slightly smaller than the canonical value of a neutron star
(1.4$M_\odot$).  We note that an inclination $\le$73$^\circ$, which
would be consistent with all other observations, pushes the mass of
the neutron star in Aql\,X-1 above this canonical value.

Following Casares et~al. (2006) we decided to use the width of the
emission lines to estimate a lower limit to the rotational
broadening, $V_{\rm rot}$$\sin$$i$, of the donor star. If we assume
a synchronously rotating companion, this gives an estimate of $q$
and $K_2$ via $V_{\rm rot}$$\sin$$i$=0.462$K_2$$q^{1/3}$$(1+q)^{2/3}$
(Wade \& Horne 1988). In order to estimate $V_{\rm rot}$$\sin$$i$ we
created an average spectrum in the rest frame of the narrow
components, and the He\,II $\lambda$4686 and H$\beta$ regions are
shown in Fig.\,\ref{rest}. From this spectrum we estimated the {\it
FWHM} of the N\,III $\lambda$4634.12 line at 147.8$\pm$36.7 km
s$^{-1}$ (we did not use N\,III $\lambda$4640.64 since it is blended
with N\,III $\lambda$4641.85). We also estimated the {\it FWHM} of
N\,III $\lambda$4634.12 in all individual spectra and find an average
of 135.7$\pm$14.8 km s$^{-1}$. We do note that there was no detectable
systematic change in {\it FWHM} as a function of orbital
phase. However, these values do include the effect of the intrinsic
broadening due to the instrumental resolution of 70 km s$^{-1}$. To
take this into account we broadened a strong line in our arc spectrum
using a Gray rotational spectrum (Gray 1992) until we reached the
observed {\it FWHM}. We assumed no limb-darking since the fluorescence
lines occur in optically thin conditions. We found that a rotational
broadening of 96$\pm$15 km s$^{-1}$ is required.

We have to be very careful with our estimate for the rotational
broadening. First of all, we assumed that all the emission of the
narrow component is coming from the companion. However, we do
note a variation in line strength (see Fig.\,\ref{trail}) and there
might also be other broadening mechanisms, as was most likely the case
in GX\,339$-$4 (Hynes et~al. 2003a). In this source the widths of
emission lines were too high (270-350 km s$^{-1}$) to be due to
rotational broadening only, and we cannot exclude that some other
broadening mechanisms are also present in Aql\,X-1. Furthermore, we
also should take into account the effect of smearing due to the
orbital motion on the emission lines. We estimated that it should be
at maximum $\simeq$20 km s$^{-1}$ for Aql\,X-1, and therefore smeared
our arc-line by this amount. We noticed that this changed our
estimates on the rotational broadening by $\le$0.6 km s$^{-1}$, small
enough to be discarded.  Finally, since the rotational broadening is
close to the spectral resolution (70 km s$^{-1}$), the optimal $V_{\rm
rot}$$\sin$$i$ found could be very sensitive to the assumed
template. We therefore also used the [O\,I] $\lambda$557.4 sky-line of
all the individual spectra to determine the rotational broadening in
this way. We notice that the variation in seeing has a small effect on
the estimate for the broadening, but it was always $\le$2 km
s$^{-1}$. Although the difference between the sky-line and arc-line is
negligible, high resolution spectra of a $\simeq$full orbit will be
needed to unambiguously determine the rotational broadening. Bearing
in mind these caveats, let us investigate the implications of our
$V_{\rm rot}$$\sin$$i$ estimates on the neutron star mass.

Combining $V_{\rm rot}$$\sin$$i$ of 96$\pm$15 km s$^{-1}$ with the
so-called $K$-correction (that allows us to estimate the centre of
mass velocity of the secondary using $K_{\rm em}$) we can determine
both $q$ and $K_2$ (Mu\~noz-Darias et~al. 2005). However, we do note
that this $K$-correction is (weakly) dependent on the inclination of
the system, but strongly dependent on the opening angle of the
accretion disk, both of which are unknown. We used the fourth order
polynomial fits given by Mu\~noz-Darias (2005) to determine the
$K$-correction (and also the corrected mass ratio $q$) that gives the
lowest estimate for the neutron star mass. We did notice that this is
a monotonically decreasing function with opening angle, i.e. the
largest opening angle gives the smallest $K$-correction and therefore
the smallest neutron star mass. On the other hand, the largest disk
opening angle does give the largest mass ratio $q$. Assuming that
$K_2$=247$\pm$8 km s$^{-1}$, we estimate that $q$$<$0.58 at 95\%
confidence (a disk opening angle of $\simeq$18$^\circ$) corresponding
to $M_1$$\sin^3$$i$$\ge$2.2$\pm$0.3$M_\odot$, larger than the
canonical neutron star mass. However, we do stress again that
high-resolution spectra are needed to verify this result, but also
note that thus far we have assumed no $K$-correction (and thus
underestimated $K_2$) or taken into account the inclination, both of
which will only increase the mass of the neutron star.

Let us in addition speculate that the velocity derived from
Fig.\,\ref{neutron} (68$\pm$5 km s$^{-1}$) gives a first order
estimate for the orbital motion of the neutron star. In this case
($q$=0.28) is close to that determined above (if $K_{\rm em}$ is close
to the centre of mass velocity $K_2$), and leads to a (maximum) disk
opening angle of $\alpha$$\simeq$16$^\circ$. Since Aql\,X-1 does not
show eclipses, we can follow Paczynski (1974, 1983) to estimate an
upper-limit on the inclination using
$\cos$$i$=0.46(1+$q$)$^{-1/3}$. This leads to $i$$<$65$^\circ$ and
together with $q$=0.28 gives an estimate on the neutron star
mass of $M_1$$>$2.7M$_\odot$, suggesting a rather massive neutron star.
Unfortunately, since we cannot trust the derived radial velocity
of the primary we cannot unambiguously conclude this, but if the $K_1$
velocity could be confirmed and the inclination of the system
constrained, it would then place strong limits on the equation of
state of neutron star material (e.g. Lattimer \& Prakash 2001).

While estimating the $K$-correction for the different disk opening
angles, we did notice that an angle of $\simeq$14-15$^\circ$ gives a
minimum mass for the neutron star $>$3.1$M_\odot$. This suggests that
already for opening angles slightly smaller than the maximum opening
angle derived above, the compact object in Aql\,X-1 will exceed the
maximum mass for a neutron star. Since we know that the compact object
is a neutron star, this could mean that the opening angle during the
May observations is close to the allowed maximum. If the opening angle
of the accretion disk was slightly larger during the time of our June
observations (either due to an asymmetry in the accretion disk or a
real increase), this would completely screen the donor star and no
narrow components would be observed anymore, explaining their absence
during the observations at the end of the outburst.

\section{Conclusion}

We conclude that we have detected narrow emission lines in the
Bowen region and He\,II $\lambda$4686 during an outburst of Aql\,X-1,
that we interpret as coming from irradiated side of the donor
star. If correct, this allows us to constrain its orbital
velocity for the first time. This leads to a hard mass function for
the neutron star of $f(M_1)$=1.23$\pm$0.12$M_\odot$, and for an
inclination $\le$73$^\circ$ exceeds the mass of a canonical
1.4$M_\odot$ neutron star. If we can trust the rotational broadening
that we determined, the width of the emission lines places a lower
limit on the mass of the primary of $\ge$1.6$M_\odot$ (at 95\%
confidence).  This suggests that Aql\,X-1 harbours a more massive
neutron star, warranting future observations exploiting the Bowen
fluorescence lines in order to improve the constraints on its binary
parameters.

Better phase-coverage of the entire outburst with high-resolution
spectra together with an updated ephemeris would not only
constrain the orbital velocity of the compact object, but also help in
understanding how the structure of the accretion disk evolves, and
further our understanding of the nature of the low-velocity
lines. Given the difficulty of observing Aql\,X-1 in quiescence, this
paper shows that high resolution optical spectroscopy could be a
promising way forward.

\section*{Acknowledgements}
We cordially thank the director of the European Southern Observatory
for granting Director's Discretionary Time (Obs Id 273.D-5026). Based
on data collected at the European Southern Observatory Paranal, Chile
(Obs Id 073.D-0819). We thank the anonymous referee for helpful comments
to this manuscript.  RC thanks Sergio Ilovaisky for useful
discussions, and Tony Bird for creating an INTEGRAL lightcurve of
Aql\,X-1. We would like to thank the ASM/RXTE teams at MIT and GSFC
for provision of the on-line ASM data. We acknowledge the use of MOLLY
and DOPPLER developed by T.R. Marsh. RC acknowledges a Marie Curie
Intra-European fellowship (MEIF-CT.2005-024685). JC acknowledges
support from the Spanish Ministry of Science and Technology through
the project AYA2002-03570. DS acknowledges a Smithsonian Astrophysical
Observatory Clay Fellowship as well as support through NASA GO grant
NNG06GC05G.

\bsp

\label{lastpage}

\end{document}